\documentclass[apjl]{emulateapj}
\usepackage{apjfonts}
\newcommand{\actaa}{Acta Astronomica}

\begin{document}
\shortauthors{Fish}
\shorttitle{OH Masers in G11.904$-$0.141}
\title{OH Masers in G11.904$-$0.141}
\author{Vincent L.~Fish}
\affil{Jansky Fellow, National Radio Astronomy Observatory, P.~O.~Box
  O, 1003 Lopezville Rd., Socorro, NM 87801, vfish@nrao.edu}
\begin{abstract}
The massive star-forming region G11.904$-$0.141 is one of only 11
sources to show maser emission in the highly-excited 13441~MHz
transition of OH.  VLBA observations of the 1665, 1667, 4765, and
13441~MHz transitions of OH toward G11.904$-$0.141 are presented.
Masers are detected at 1665, 1667, and 4765~MHz, but the 13441~MHz
masers are not detected.  Consistent magnetic field strengths of
approximately $+3.5$~mG are detected in the ground-state masers, in
contrast with a possible $-3.0$~mG magnetic field previously detected
at 13441~MHz.  The variable 13441~MHz masers may be associated with an
outflow.
\end{abstract}

\section{Introduction}

Massive star-forming regions in the Galaxy are commonly seen to host
OH maser emission, including masers from excited states.  However,
maser emission in the highly-excited 13441~MHz transition is rare.  To
date, only 11 sources are known to host 13441~MHz masers, most
detected only recently \citep{turner70,baudry02,caswell04}.  In
addition, several possible sources were detected by \citet{balister76}
but not subsequently redetected.

Nearly all 13441~MHz maser sources display nearly equal fluxes in the
left (LCP) and right circular polarized (RCP) modes.  Conventional
wisdom attributes this to the small Zeeman splitting coefficient
(0.018~km\,s$^{-1}$~mG$^{-1}$).  Since a typical magnetic field of a
few milligauss does not split the LCP and RCP lines of a Zeeman pair
by a full linewidth, velocity-coherent amplification of the LCP and
RCP lines should be similar, giving a flux ratio of nearly unity.
Approximately equal LCP and RCP fluxes are observed in 10 of 11
13441~MHz OH maser sources, including for individual Zeeman pairs
observed at very long baseline interferometric (VLBI) resolution in
\object{W3(OH)} \citep{baudry98}.  The one exception is
\object[G11.904-0.141]{G11.904$-$0.141} (hereafter G11.90), for which
the RCP/LCP flux ratio was 3.4 in the \citet{caswell04} observations
and larger in the \citet{baudry02} observations, in which the LCP
feature was not detected.  \citet{caswell04} infers a possible
magnetic field of $-3.0$~mG, which produces a negligible splitting
compared to the line width (0.24~km\,s$^{-1}$ RCP and
0.31~km\,s$^{-1}$ LCP) and therefore would be expected to result in
fairly equal RCP and LCP fluxes.

Observations of multiple OH maser transitions at VLBI resolution would
also provide information on whether the several maser transitions are
co-spatial, with implications for 13441~MHz maser modelling.
Multitransition maser overlaps provide strong observational
constraints to test maser models and to derive local physical
conditions.  The literature includes only one source, W3(OH), observed
at 13441~MHz at VLBI resolution \citep{baudry98}.  Further
observations are needed in order to understand the conditions that
produce these rare masers.

It is for these reasons that G11.90 was selected for study at higher
spatial resolution.  Results are reported in this Letter.

\section{Observations}

Four transitions of OH masers were observed with the Very Long
Baseline Array (VLBA) in three different epochs (experiment code
BF088).  The ground-state 1665.40184 and 1667.35903~MHz masers were
observed simultaneously on 2006~Feb~22 and the 4765.562~MHz masers on
2006~Feb~27.  Total observing time was 6.5~hr per run, with
approximately 3~hr spent on G11.90.  The 13441.4173~MHz masers were
observed with both the VLBA and the Green Bank Telescope (GBT) on
2006~Aug~29 over a total of 4.5~hr, with approximately 2~hr spent on
G11.90.  Most of the remaining time was used to observe the nearby
calibrator J1825$-$1718 for phase calibration.  The calibrator 3C286
was observed as a fringe-finder and polarization calibrator.

The 1665, 1667, and 4765~MHz masers were observed in full polarization
mode using 0.125~MHz bandwidths divided into 128 spectral channels for
a spectral resolution of 0.18~km\,s$^{-1}$ at 1665/1667~MHz and
0.06~km\,s$^{-1}$ at 4765 MHz.  The 13441~MHz masers were observed in
dual circular polarization mode using a 1.0~MHz bandwidth divided into
512 spectral channels for a spectral resolution of 0.04~km\,s$^{-1}$.

Because G11.90 is significantly scatter-broadened at long wavelengths,
no signal was detectable on the longer baselines at 1.6~GHz.  The
usable array at 1.6~GHz effectively consisted of the inner five
antennas plus a small amount of data from North Liberty, IA.  Blank
sky noise in the LCP and RCP image cubes was 18~mJy\,beam$^{-1}$ in a
single channel with a synthesized beam size of $67 \times 32$~mas.  At
4.7~GHz, sufficient signal was seen to determine adequate calibration
on baselines to all antennas except Mauna Kea, HI and St.~Croix, VI.
Blank sky noise in Stokes I was 8~mJy\,beam$^{-1}$ in a $9 \times
3$~mas synthesized beam.  At 13441~GHz all 10 VLBA antennas and the
GBT produced usable data, for a blank sky Stokes I noise level of
3~mJy\,beam$^{-1}$ in a $2.8 \times 0.7$~mas beam when averaged over 5
spectral channels, comparable to a maser line width.

All data were phase-referenced to J1825$-$1718 ($3\fdg5$ away from
G11.90) using a cycle time of 5~min at 1665/1667~MHz and 3~min at 4765
and 13441~MHz.  The phase-referenced 1665 and 1667~MHz image quality
was poor, but the reference feature (1665~MHz LCP) was sufficiently
bright to determine a position.  The reference maser was then used to
self-calibrate the 1665 and 1667~MHz LCP and RCP data for further
imaging.  Phase-referencing at 4765~MHz produced excellent image
quality.  Random errors from centroid fitting in determining reference
feature positions are less than 1~mas at 1665 and 4765~MHz.  However,
systematic errors can dominate at 1.6~GHz, with ionospheric
fluctuations able to introduce apparent shifts on the order of 20~mas
as well as ambiguities of which peak corresponds to the actual source
for nodding calibration under poor ionospheric conditions
\citep{chatterjee99}.  Much of the power in the phase-referenced
1665~MHz map was distributed among sidelobes, but there was a clear
brightest feature, with the next-brightest sidelobes distributed
symmetrically about this feature.  It therefore appears that the
ionosphere was sufficiently stable during observations to permit
phase-referencing to successfully determine the position of the
1665~MHz LCP feature, with likely errors no more than a few
milliarcseconds.

\begin{deluxetable*}{lcllllrrrc}
\tablecaption{Detected Masers in G11.90 \label{masers}}
\tablehead{
  \colhead{Freq.} &
  \colhead{} &
  \colhead{RA\tablenotemark{a}} &
  \colhead{Decl.\tablenotemark{a}} &
  \colhead{$v_\mathrm{LSR}$\tablenotemark{a}} &
  \colhead{$S$\tablenotemark{a}} &
  \colhead{} &
  \colhead{$m_L$\tablenotemark{c}} & 
  \colhead{$\chi$\tablenotemark{c}} &
  \colhead{$B$} \\
  \colhead{(MHz)} &
  \colhead{Pol.} &
  \colhead{(J2000)} &
  \colhead{(J2000)} &
  \colhead{(km\,s$^{-1}$)} &
  \colhead{(Jy)} &
  \colhead{$N_{\mathrm{chan}}$\tablenotemark{b}} &
  \colhead{(\%)} &
  \colhead{(\degr)} &
  \colhead{(mG)} 
}
\startdata
1665 & LCP & 18~12~11.4314 & $-$18~41~28.827                 & 40.22 & 1.33 & 3&     8 & $-$11 & 3.6   \\ 
1665 & RCP & 18~12~11.4319 & $-$18~41~28.827                 & 42.33 & 0.22 & 1&\nodata&\nodata&       \\ 
1665 & LCP & 18~12~11.4338 & $-$18~41~28.803\tablenotemark{d}& 40.75 & 9.28 & 5&    10 & $-$43 & 3.6   \\ 
1665 & RCP & 18~12~11.4333 & $-$18~41~28.807                 & 42.85 & 0.84 & 2&\nodata&\nodata&       \\ 
1665 & LCP & 18~12~11.4352 & $-$18~41~28.793                 & 41.62 & 0.44 & 3&\nodata&\nodata&\nodata\\ 
1667 & LCP & 18~12~11.4316 & $-$18~41~28.827                 & 40.76 & 0.22 & 1&\nodata&\nodata&\nodata\\ 
1667 & LCP & 18~12~11.4334 & $-$18~41~28.810                 & 41.11 & 1.00 & 3&    13 &     0 & 3.5   \\ 
1667 & RCP & 18~12~11.4332 & $-$18~41~28.809                 & 42.34 & 0.41 & 2&\nodata&\nodata&       \\ 
4765 & I   & 18~12~11.4368 & $-$18~41~28.883                 & 41.86 & 0.45 & 3&    25\tablenotemark{e}& 70 &\nodata\\
4765 & I   & 18~12~11.4371 & $-$18~41~28.881                 & 41.92 & 1.09 & 5&     5 &    88 &\nodata  
\enddata
\tablenotetext{a}{{}Position, central velocity, and flux density of channel of peak emission.}
\tablenotetext{b}{{}Number of channels in which feature is detected.}
\tablenotetext{c}{{}Linear polarization fraction and (electric vector) polarization angle east of north.}
\tablenotetext{d}{{}Reference feature for ground-state masers.}
\tablenotetext{e}{{}Some contamination from nearby brighter feature.}
\end{deluxetable*}

\section{Results}

Detected masers are detailed in Table~\ref{masers}.  Flux
densities are given for the peak channel of emission for each spot in
the natural polarization mode: LCP and RCP for the 1665 and 1667~MHz
transitions (for which the Zeeman splitting coefficient is large) and
Stokes I for the 4765~MHz transition (in which the Zeeman splitting
coefficient is a factor of $\sim 1000$ smaller).  No masers were
detected at 13441~MHz.

It is possible that several maser spots are spatially blended together
in the ground state.  At a distance of 5.1~kpc \citep{caswell95}, the
synthesized beam size corresponds to a spatial size of over 200~AU,
which is larger than the typical OH maser clustering scale
\citep{fish06}.  The ground-state masers are significantly broadened,
although not atypically so for sources at low Galactic longitude
\citep[e.g.,][]{szymczak85,fish06}, while the 4765~MHz masers appear
to be much smaller, consistent with interstellar scattering varying
as $\nu^{-2}$ or $\nu^{-2.2}$ \citep[e.g.,][]{cordes84}.

Emission at 1665 and 1667~MHz is coincident to a small fraction of the
beam size.  Where 1665 and 1667~MHz masers overlap, magnetic field
strengths and systemic velocities (corrected for the Zeeman effect)
are in excellent agreement.  Masers in the 4765~MHz transition are
offset slightly to the southeast from the ground-state masers though
at approximately the same LSR velocity as ground-state masers when the
latter are corrected for Zeeman splitting.  All OH masers fall in the
velocity range spanned by 6668 and 12178~MHz methanol masers
\citep{menten91,caswell95a,caswell95b,szymczak00}.  The total spatial
extent of the 1665, 1667, and 4765~MHz masers is less than 100~mas
(500~AU) in each of the right ascension and declination directions.
Maser emission in the 6035~MHz transition spans 130~mas, mostly in the
declination direction \citep{caswell97}.  The positions of the
northern and southern 6035~MHz masers are consistent with being
approximately coincident with the line of ground-state masers and
4765~MHz maser group, respectively, given the positional uncertainty
of the \citet{caswell97} Australia Telescope Compact Array (ATCA)
observations,

All masers brighter than 1~Jy are seen in linear polarization as well,
typically at the 10\% level.  The weaker 4765~MHz spot appears to have
a large linear polarization fraction but may suffer from contamination
from the nearby brighter spot.  The polarization position angles are
likely affected by large Faraday rotation between the source and the
observer, since the highly-broadened maser sizes at 1.6~GHz imply a
large electron density along the propagation path.  Typical dispersion
measures in the general direction and near the distance of G11.90 are
250--500~cm$^{-3}$\,pc \citep[from the ATNF Pulsar
Catalogue,][]{manchester05}.  Assuming an average line-of-sight
Galactic magnetic field of 2~$\mu$G \citep[e.g.,][]{heiles76},
expected rotation measures imply a Faraday rotation of greater than
1~rad at 4765~MHz and tens of radians at 1.6~GHz.

No emission is detected at 13441~MHz, to a $5\,\sigma$ limit of
20~mJy\,beam$^{-1}$ in each of LCP and RCP when the spectral
resolution is degraded to 0.22~km\,s$^{-1}$, comparable to the maser
line width in the \citet{caswell04} detection.  RCP emission was
detected at the $200$~mJy level in 1999 May by \citet{baudry02}, while
LCP emission was not detected at all.  RCP and LCP flux densities were
measured to be 240~mJy and 70~mJy, respectively, by \citet{caswell04}
in 2003 June.  The 13441~MHz masers in G11.90 are clearly variable.
Despite the small number of known 13441~MHz maser sources and the
relative paucity of observations thereof compared to less-excited
states, 13441~MHz masers are known to display large variability in
general \citep[see the discussion in][]{caswell04}.  Variability has
been noted in other maser transitions in G11.90 as well, including OH
6035~MHz \citep{caswell95} and methanol 6668~MHz \citep{szymczak00}.
In the other transitions reported herein, the 1665~MHz masers have
become moderately stronger and the 1667~MHz masers moderately weaker
since the \citet{caswell83} epoch (1980--1981), while the 4765~MHz
maser parameters (peak velocity, flux density, and position) are all
in excellent agreement with the ATCA observations of \citet[][epoch
2000 September 15--16]{dodson02}.  A monitoring study of the 4765~MHz
masers show no significant variability over nearly two years
\citep{smits03}.

Magnetic fields of about $+3.5$~mG are detected in the ground-state
transitions, where the positive sign represents a magnetic field whose
line-of-sight component is oriented away from the observer.  LCP and
RCP velocities were determined by taking the central velocity of peak
emission; corresponding magnetic field strength errors are 0.2~mG at
1665~MHz and 0.4~mG at 1667~MHz.  These results are consistent with
the Parkes spectra of \citet{caswell83}, which show RCP emission
redshifted with respect to stronger LCP emission at both 1665 and
1667~MHz.  However, the positive magnetic field values contrast with
observations in the excited states.  \citet{caswell95} detected no
Zeeman splitting at 6035~MHz, placing an upper limit of $|B| <
0.5$~mG.  At 13441~MHz, \citet{caswell04} detected a marginal magnetic
field of $-3.0$~mG, of similar strength to the detections reported
herein but opposite in sign.  The 13441~MHz splitting may also be
consistent with a zero magnetic field, as indicated by 6035~MHz Zeeman
splitting, but is not consistent with the $+3.5$~mG magnetic fields
obtained in the ground-state transitions.  It is possible that there
is a magnetic field direction reversal across the source, but the VLBA
nondetection of the 13441~MHz masers precludes definitive conclusions
regarding their location relative to the ground-state masers.

\section{Discussion}

All masers are located near continuum source A in \citet{forster00}.
At 8.2 and 9.2~GHz, the continuum emission appears to have two peaks,
with the brighter, more compact peak,
\object[G11.904-0.142]{G11.904$-$0.142}, located nearest the OH
emission, to the northeast of the weaker, more extended peak
\citep{forster00}.  Extended emission oriented northeast-southwest
connects the two peaks, with larger-scale weaker emission elongated
north-south.  The 1665 abd 1667~MHz masers are distributed as 3
distinct sites of emission along a line oriented at a position angle
of 55\degr\ (east of north), with velocities increasing toward the
northeast.  The absolute position of the brightest peak is located
approximately 880~mas to the northwest from the location given in
\citet{forster99}, which is reasonably consistent to within the errors
that the authors quote.

The maser distribution and velocities hint at a possible outflow
origin.  Assuming that the 13441~MHz masers, when they are detectable,
exist near the other masers, they too may be associated with the
putative outflow.  If the negative magnetic field marginally detected
at 13441~MHz can be believed \citep{caswell04}, it is possible that
the magnetic field wraps around the outflow, with the 13441~MHz masers
spatially offset from the ground-state masers.  Should the 13441~MHz
masers reappear and become sufficiently bright, further
interferometric study would be useful to test this hypothesis.  Linear
polarization measurements of the 13441~MHz masers would also allow the
direction of the magnetic field in the plane of the sky to be
inferred.

If an outflow does exist in G11.90, it may be partially responsible
for powering the 13441~MHz masers.  In other sources, the association
of some masers with outflows is clearest in methanol \citep{zapata06}
but is seen in other tracers as well, including OH
\citep[e.g.][]{debuizer06}.  OH transitions in star-forming regions
are inverted via a complex system of radiative and collisional pumping
routes whose relative importance is likely dependent on physical
conditions in the maser region \citep{gray07}.  Pumping of the
13441~MHz transition is not well understood but probably requires
quite energetic conditions, given the rarity of 13441~MHz masers as
well as the large excitation of the $^2\Pi_{3/2}, J = 7/2$ state
(290~K above ground).  Indeed, in W3(OH), the only source in which
13441~MHz masers have been detected at VLBI resolution, the 13441~MHz
masers are located in a region of high excitation where the methanol
maser kinematics suggest the existence of an expanding bipolar cone
\citep{baudry93,baudry98,moscadelli02}.  Association of 13441~MHz
masers with energetic outflows may also naturally explain the high
variability usually seen in this transition \citep{caswell04}.
Further high-resolution studies of 13441~MHz masers, as well as
contectualizing high-resolution information from molecular tracers and
infrared emission, are required to understand their excitation and
their relation to their surroundings.

\acknowledgments

The National Radio Astronomy Observatory is a facility of the National
Science Foundation operated under cooperative agreement by Associated
Universities, Inc.  The ATNF Pulsar Catalogue can be found at
\url{http://www.atnf.csiro.au/research/pulsar/psrcat}~.

{\it Facilities: \facility{VLBA}, \facility{GBT}}

\end{document}